\documentclass[prd,10pt,aps,twocolumn,nofootinbib,superscriptaddress,showpacs,floatfix,amssymb,amsmath]{revtex4-1}
\usepackage{rotating}

\newcommand{\ext}{{\mathrm{ext}}}

\newcommand{\beqn}{\begin{eqnarray}}
\newcommand{\eeqn}{\end{eqnarray}}
\newcommand{\eq}[1]{(\ref{#1})}

\newcommand{\cL}{{\cal L}}

\newcommand{\EW}{{\mathrm{EW}}}

\newcommand{\Z}{{Z \!\!\! Z}}

\newcommand{\be}{\begin{equation}}
\newcommand{\ee}{\end{equation}}
\newcommand{\bal}{\begin{align}}
\newcommand{\eal}{\end{align}}
\newcommand{\bmt}{\begin{multline}}
\newcommand{\emt}{\end{multline}}

\begin{document}

\title{Magnetic-field-induced superconductivity and superfluidity of $\boldsymbol W$ and $\boldsymbol Z$ bosons: \\ In tandem transport and kaleidoscopic vortex states}

\author{M. N. Chernodub}\thanks{On leave from ITEP, Moscow, Russia.}
\affiliation{CNRS, Laboratoire de Math\'ematiques et Physique Th\'eorique, Universit\'e Fran\c{c}ois-Rabelais Tours,\\ F\'ed\'eration Denis Poisson, Parc de Grandmont, 37200 Tours, France}
\affiliation{Department of Physics and Astronomy, University of Gent, Krijgslaan 281, S9, B-9000 Gent, Belgium}
\author{Jos Van Doorsselaere}
\affiliation{Department of Physics and Astronomy, University of Gent, Krijgslaan 281, S9, B-9000 Gent, Belgium}
\author{Henri Verschelde}
\affiliation{Department of Physics and Astronomy, University of Gent, Krijgslaan 281, S9, B-9000 Gent, Belgium}

\begin{abstract}
We show that in a background of a sufficiently strong magnetic field the electroweak sector of the quantum vacuum exhibits superconducting and, unexpectedly, superfluid properties due to the magnetic-field-induced condensation of, respectively, $W$ and $Z$ bosons. The phase transition to the ``tandem'' superconductor--superfluid phase -- which is weakly sensitive to the Higgs sector of the standard model -- occurs at the critical magnetic field of $10^{20}$~T. The superconductor-superfluid phase of the electroweak vacuum has anisotropic transport properties as both charged and neutral superflows may propagate only along the magnetic field axis. The ground state possesses an unusual ``kaleidoscopic'' structure made of a hexagonal lattice of superfluid vortices superimposed on a triangular lattice of superconductor vortices. 
\end{abstract}


\date{\today}

\maketitle

\section{Introduction}

		It is known that an extremely high magnetic field of hadronic scale may lead to plenty of unusual effects both in (dense) matter and in the quantum vacuum. The chiral magnetic effect~\cite{Fukushima:2008xe} provides a particularly interesting example: charge-parity-odd matter may generate an electric current along the axis of the magnetic field~\cite{Vilenkin:1980fu}. The corresponding conditions may be realized in noncentral heavy-ion collisions~\cite{Kharzeev:2004ey} in which a hot quark matter is created along with a background of extremely high magnetic fields~\cite{Skokov:2009qp,Deng:2012}. Similar conditions may have existed in the very early moments of our Universe~\cite{Grasso:2000wj}. The strong magnetic field also affects phases of the cold dense matter in the cores of strongly magnetized neutron stars~\cite{Ferrer:2005vd}.
		
		Because of quantum effects an empty space may also exhibit quite unusual properties in a sufficiently strong magnetic  background. In the background of a relatively low magnetic field of QED scale the vacuum should become optically birefringent~\cite{Adler:1971wn}.  The hadron-scale magnetic field should lead to magnetic catalysis~\cite{Klimenko:1991he}, which implies, in particular, a steady enhancement of the chiral symmetry breaking in QCD vacuum as the external magnetic field strengthens. 
		
		More recently it was found that the vacuum becomes an {\it electromagnetic} superconductor in sufficiently strong external magnetic fields~\cite{Chernodub:2010qx,Chernodub:2011mc}. The superconductivity of, basically, empty space, is mediated via spontaneous creation of a (charged) $\rho$-meson condensate if the magnetic field exceeds the critical value of 
$B^{\mathrm{QCD}}_c  \simeq 1.0 \times 10^{16}\,\mbox{T}$.
The ground state of the vacuum superconductor is characterized by inhomogeneous ground state of a very particular geometric structure~\cite{Chernodub:2011gs}, possessing intriguing metamaterial (``perfect lens'') properties~\cite{Smolyaninov:2011wc}. The magnetic fields of the required strength may be created on  Earth in heavy-ion collisions at the Large Hadron Collider at CERN~\cite{Deng:2012}.
		
		We show that as the background magnetic field strengthens further, the Standard Model experiences a second superconducting, and, simultaneously, superfluid transition associated with a condensation of the $W$ and $Z$ bosons at a larger critical magnetic field:
			\beqn
			B^{\mathrm{EW}}_c = \frac{M_W^2}{e} \simeq 1.1 \times 10^{20}\,\mbox{T}\,,
			\label{eq:eBc}
			\eeqn
		where $M_W = 80.4 \, \mbox{GeV}$ is the mass of the $W$ boson. The onset of the condensation of the $W$ bosons at the magnetic field~\eq{eq:eBc} was predicted by Ambj\o rn and Olesen in Ref.~\cite{Ambjorn:1988tm}. The key idea here is that the vacuum of charged vector particles (i.e., of the $W$ mesons) is unstable in the background of a sufficiently strong magnetic field provided these particles have anomalously large gyromagnetic ratio $g_{m} = 2$. The large value of $g_m$ guarantees that the magnetic moment of such particles is too large to withstand a spontaneous condensation at sufficiently strong external magnetic fields. In this article we show that the inhomogeneous $W$ condensation induces an inhomogeneous condensation of the $Z$ bosons and leads to new superconducting and superfluid effects at the electroweak scale.

The electroweak sector possesses another phase transition which lifts off the electroweak symmetry breaking at a second critical magnetic field $B^{\mathrm{EW}}_{c2}$ which is stronger than the critical magnetic field of the electroweak superconducting transition~\eq{eq:eBc}, $B^{\mathrm{EW}}_{c2} > B^{\mathrm{EW}}_{c1} \equiv B^{\mathrm{EW}}_c $~\cite{ref:second:transition:1}. A recent study of the second phase transition can be found in Ref.~\cite{ref:second:transition:2}. In this paper we concentrate on the $W$-meson condensed phase realized at $B^{\mathrm{EW}}_{c1} < B < B^{\mathrm{EW}}_{c2}$.

The structure of this paper is as follows. In Sec.~\ref{sec:structure} we solve the classical equations at the $W$-condensed phase and we show that the $W$ condensate, originally found in Ref.~\cite{Ambjorn:1988tm}, is accompanied by the condensation of the electrically neutral $Z$ bosons. We point out that both condensates possess vortex defects which are aligned with the magnetic field axis forming a complicated regular structure in the transversal plane. In Sec.~\ref{sec:transport} we demonstrate that these condensates lead to the new transport phenomena of the ground state, which correspond to a dissipationless transfer of an electric current and a neutral $Z$-boson current. We associate these phenomena with superconductivity and superfluidity, respectively. The last section is devoted to our conclusions.

\section{Structure of the ground state}
\label{sec:structure}

\subsection{Equations of motion}

		The bosonic part of the electroweak sector of the Standard Model is described by the Lagrangian,
			\beqn
			\mathcal{L} & = & -\frac{1}{4}W^a_{\mu\nu} W^{a,\mu\nu}-\frac{1}{4}X_{\mu\nu}X^{\mu\nu} + (D_\mu\Phi)^\dagger  (D^\mu\Phi) \nonumber \\
			& & -  \lambda \left(\vert\Phi\vert^2- v^2/2\right)^2\,.
			\label{eq:L}
			\eeqn
where $\Phi$ is the complex Higgs doublet which interacts with $SU(2)_L$ and $U(1)_X$ gauge fields ($W^a_\mu$ and $X_\mu$, respectively) via the covariant derivative $$D_\mu = \partial_\mu - i g \tau^a W^a_\mu/2 {-} i g' X_\mu /2\,,$$ and $\tau^a$ are the Pauli matrices. The corresponding field strengths are $W^a_{\mu\nu} = \partial_\mu W^a_\nu - \partial_\nu W^a_\mu + g \epsilon^{abc} W^b_\mu W^c_\nu$ and $X_{\mu\nu} = \partial_\mu X_\nu - \partial_\nu X_\mu$.

		The Mexican hat potential in Eq.~\eq{eq:L} breaks the electroweak symmetry down to the electromagnetic subgroup, $SU(2)_L \times U(1)_X \to U(1)_{\mathrm{em}}$ because the Higgs field $\Phi$ acquires a quantum expectation value, $\langle \Phi \rangle \neq 0$. In the unitary gauge, $\langle \Phi \rangle = (0,v)^T$, the third component of the non-Abelian gauge field $W^3_\mu$ mixes with the Abelian gauge field $X_\mu$ providing us with the massive  $Z_\mu$ boson and the massless electromagnetic field~$A_\mu$:
\beqn
W^3_\mu & = & \sin\theta A_\mu+\cos\theta Z_\mu\,, \\
X_\mu & = & \cos\theta A_\mu-\sin\theta Z_\mu\,,
\eeqn
where $\theta$ is the electroweak mixing (Weinberg) angle with $e=g \sin\theta = g' \cos\theta$ being the electric charge.
		
		The classical equations of motion are as follows:
			\beqn
			0 & = & \partial^\mu W_{\mu\nu}^a+g\epsilon^{abc}W^{b\mu}W^c_{\mu\nu} {-} i g  \left[(D_\mu\Phi)^\dag \tau^a \Phi {-} h.c.\right]/2\,, \nonumber\\
			0 & = & \partial^\mu X_{\mu\nu} - i g' \left(D_\mu\Phi)^\dag  \Phi-h.c.\right)/2\,, 
			\label{eq:classical}\\
			0 & = & -D_\mu D^\mu\Phi+ 2 \lambda \, \Phi(\vert\Phi\vert^2-v^2/2) \nonumber\,,
			\eeqn

\begin{figure}[!ht]
\begin{center}
\includegraphics[scale=0.45,clip=false]{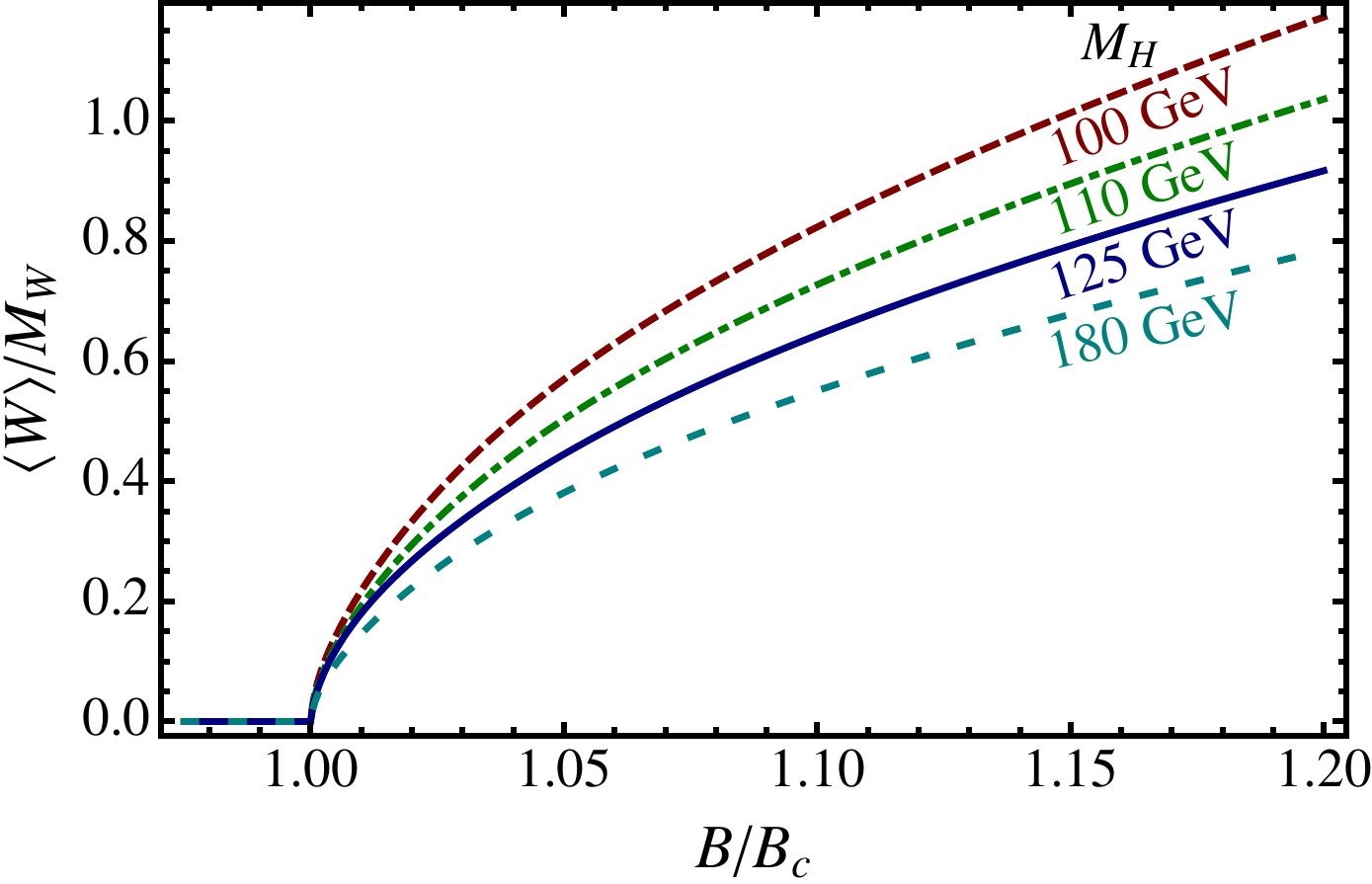} \\[1mm] 
\hskip 6mm (a) \\[3mm]
\hskip -6mm \includegraphics[scale=0.5,clip=false]{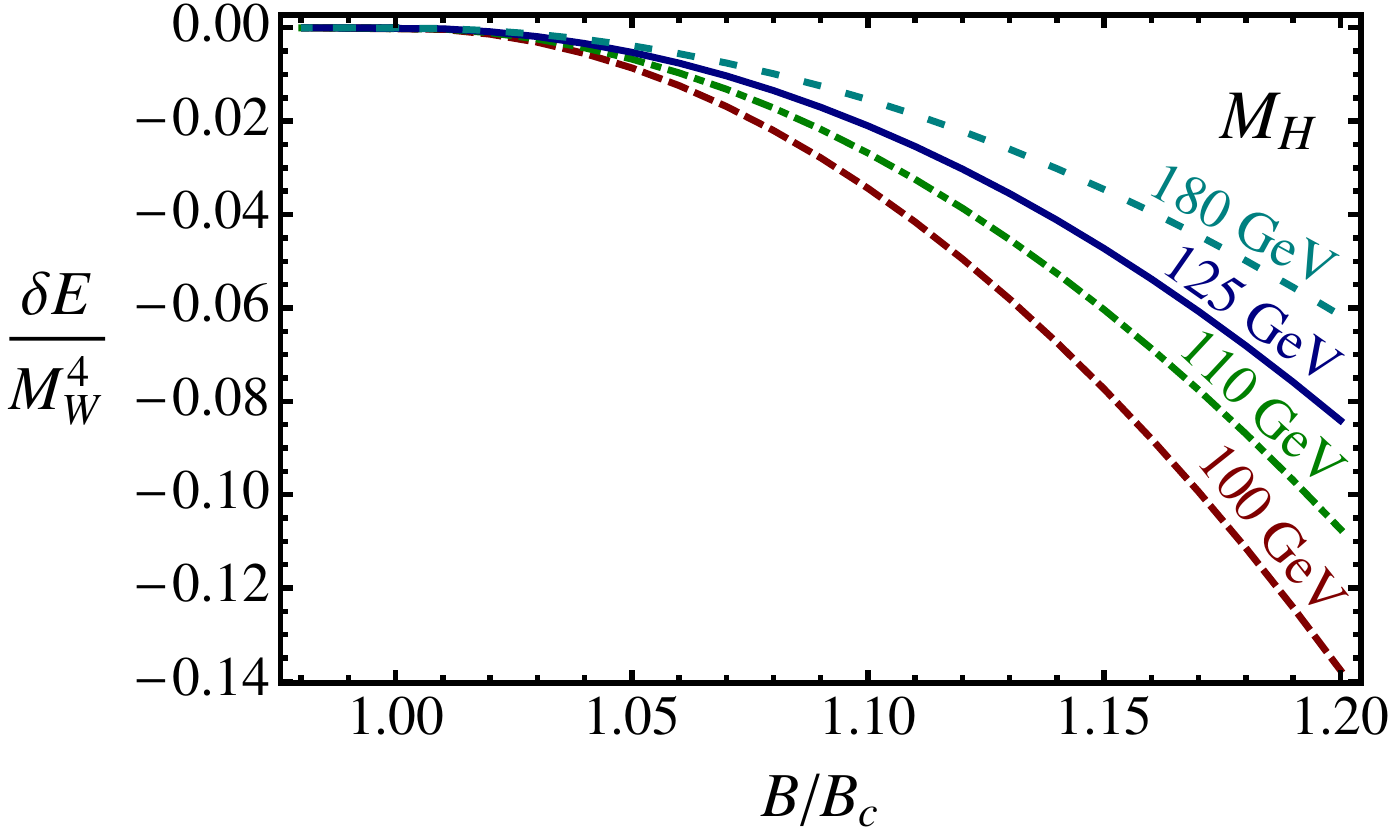} \\[1mm] 
\hskip 6mm (b) \\[3mm]
\includegraphics[scale=0.45,clip=false]{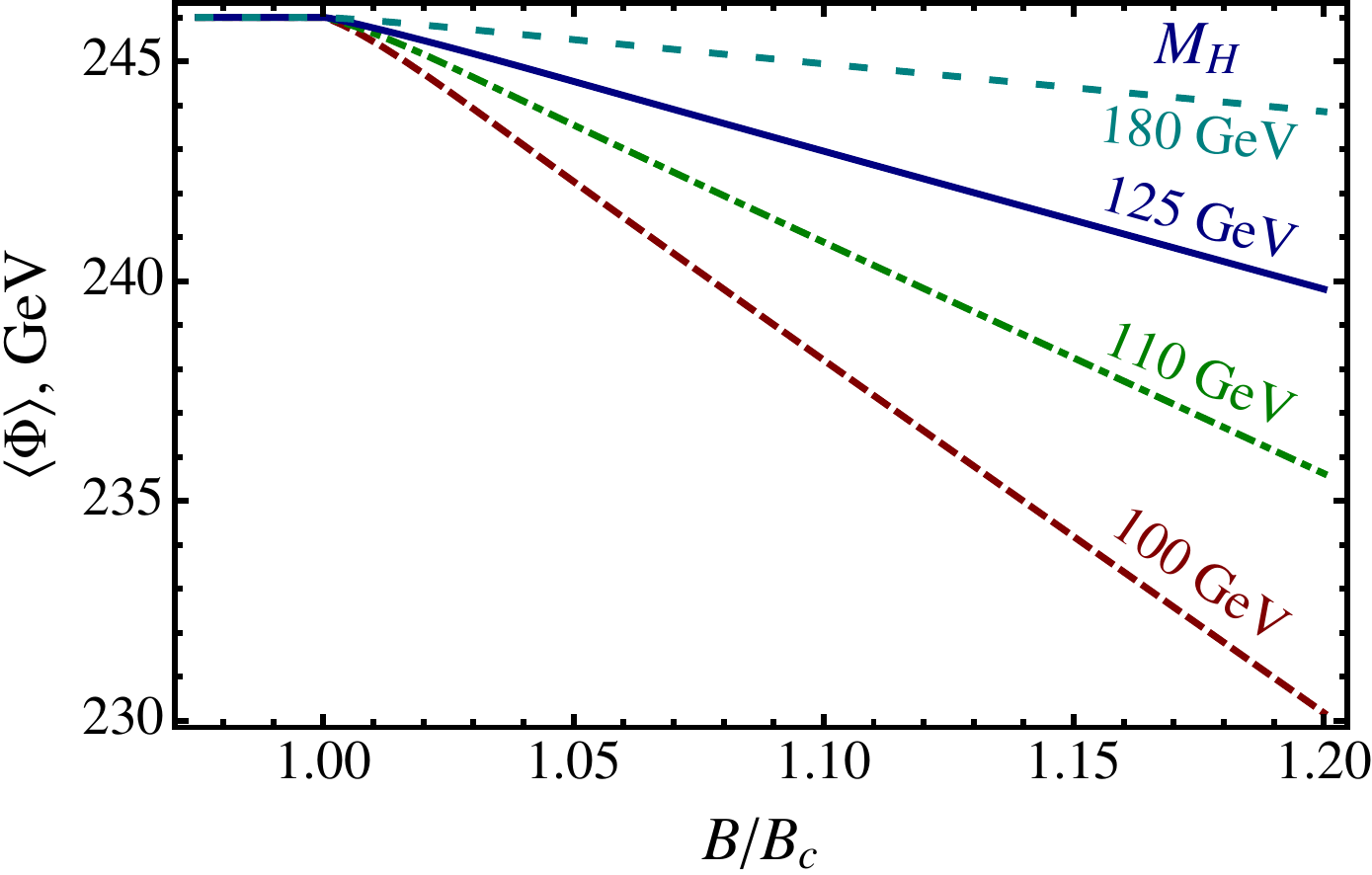} \\[1mm] 
\hskip 6mm (c)
\end{center}
\caption{(a) The cell-averaged $W$ condensate $\langle |W|^{2}\rangle^{1/2}$; (b) the condensation energy density~\eq{eq:E}, $\delta E = \langle E \rangle_{W} - \langle E \rangle_{W=0}$ and (c) the cell-averaged Higgs expectation value $\langle \Phi^{\dagger} \Phi\rangle^{1/2}$, Eq.~\eq{eq:phi}, vs. the strength of the magnetic field $B$ (in units of the critical magnetic field $B \equiv B^{\EW}_{c}$). The plots are given for various Higgs masses $M_H$ including the physical value of the Higgs mass (the latter are shown by the solid lines).}
\label{fig:energy}
\end{figure}
	
		The instability of the vacuum in the presence of the sufficiently strong magnetic field was first demonstrated by Ambj\o rn and Olesen in Ref.~\cite{Ambjorn:1988tm}, and we briefly repeat their arguments here. We restrict ourselves to the classical dynamics of the electroweak fields and ignore quantum corrections following the original approach of Ref.~\cite{Ambjorn:1988tm}, which is justified in a sufficiently strong classical background.
	
Consider a uniform time-in\-de\-pen\-dent magnetic field directed along the third axis, $B_{\ext,i} = B_\ext \delta_{i3}$ (for the sake of convenience, we always take $eB_\ext >0$). Then the quadratic part of the transverse (with respect to the magnetic field axis) components of the $W_\mu \equiv W^-_\mu$ field in Eq.~\eq{eq:L} reads
			\beqn
			\delta {\mathcal L}^{(2)}_{W_\perp} = 
			\left( W_1^\dagger, W_2^\dagger\right)
			\left(
				\begin{array}{cc}
				M_W^2 & - i e B_\ext\\
				i e B_\ext & M_W^2 
				\end{array}
			\right)
			\left( 
				\begin{array}{c}
				W_1\\
				W_2
				\end{array}
			\right)\,, \quad
			\label{eq:W2}
			\eeqn
		(with $M_W = g v/2$), while the mass terms of the longitudinal components $W_{3}$ and $W_0$, and of other vector particles are not affected at the classical level. The mass eigenvalues of Eq.~\eq{eq:W2} are $\mu^2_\pm = M^2_W \pm eB$. One of the masses, $\mu_-$, vanishes at the critical value $B_c$ of the magnetic field~\eq{eq:eBc}. This mass becomes purely imaginary at $B > B_c$, thus signaling a tachyonic instability towards condensation of the transverse components of the $W_{\mu}$ field. The unstable eigenvector is $(W_1,W_2) = (W, - i W)/2$, where $W$ is a scalar field.

		Since we consider the solutions in the transverse $(x_1,x_2)$ plane for the transverse components of the fields, it is natural 
to use the complex notation for the coordinates, $z=x_1+ix_2$, and for the vectors $\mathcal{O}_\mu = \partial_\mu$, $A_\mu$, $Z_\mu$, $W_\mu$: $\mathcal{O}=\mathcal{O}_1+i\mathcal{O}_2,\ \mathcal{\bar O} = \mathcal{O}_1 - i\mathcal{O}_2$, and their field strengths: $ {\mathcal O}_{12}= - \frac{i}{2} (\bar\partial {\mathcal O} - \partial {\mathcal {\bar O}})$. Notice that $W^\dagger \neq {\bar W}$. 
		
		We use a symmetric gauge for the external magnetic field, $A_{\ext,1} = - B x_2/2$ and $A_{\ext,2} = B x_1/2$, so that the corresponding covariant derivative is $$\mathcal{D}_\ext \equiv \partial-i e A_\ext=\partial+e z B_\ext /2\,.$$ The $\bar W$ component of the $W^-$ corresponds to the $\mu_+$ eigenvalue of the operator in Eq.~\eq{eq:W2}, so that it is not condensed. Thus, we put $\bar W=0$ as it does not lower the energy, and 
		continue to 
		work in the unitary gauge $\Phi = (0,\phi)^T$, where $\phi$ is a real-valued field.
		
		In complex notation the energy density is
			\beqn
			E & = & \frac{1}{2}\left\vert(\mathcal{\bar D}+ig\cos\theta \bar Z)W\right\vert^2+\frac{1}{2}Z_{12}^2+\frac{1}{2}B^2+\frac{g^2}{8}\vert W\vert^4\nonumber \\
			& & \!\!\! + \frac{1}{2}(-e  B-g\cos\theta  Z_{12} +\frac{g^2}{2}\phi^2)\vert W\vert^2 + \frac{g^2}{4\cos^2\theta}\vert Z\vert^2 \phi^2 \nonumber \\
			& & \!\!\! + \bar\partial\phi \partial\phi + \lambda(\phi - v^2/2)^2\,,
			\label{eq:E}
			\eeqn
		and the equations of motion then become as follows:
			\beqn
			\mathfrak{D}\mathfrak{\bar D}W&=&  \left( \frac{g^2}{2} \vert W\vert^2 - g\cos\theta Z_{12}- eF_{12} + \frac{g^2}{2} \phi^2\right)W \,,
			\label{eq:EoM1} 
			\\
			   \mathfrak{\bar D}^2W& = &0 \,, 
			\label{eq:EoM2} 
			\\
			  \bar \partial F_{12}&=& \frac{e}{2}\bar\partial \vert W\vert^2 + \frac{e}{2} W^\dag\mathfrak{\bar D}W ,
			\label{eq:EoM3} 
			\\
			0&=& \cos\theta\bar\partial F_{12}-\sin\theta \bar\partial Z_{12}+ig^2\frac{\sin\theta}{2\cos^2\theta}\bar Z\phi^2 
			\label{eq:EoM4} \\
			 \partial\bar\partial\phi & = & \frac{g^2}{4\cos^2\theta}\vert Z\vert^2 \phi+ \frac{g^2}{4}\vert W\vert^2\phi + 2 \lambda \phi (\phi^2- \frac{v^2}{2}),\qquad
			\label{eq:EoM5} \\
			0 & = & \phi \mathfrak{\bar D}W+2W\left(\bar\partial+i \frac{g}{2\cos\theta}\bar Z\right)\phi\,,
			\label{eq:EoM6} 
			\eeqn
		where $\mathfrak D = \mathcal{D}+i g\cos\theta Z$ is a covariant derivative.
		
In Ref.~\cite{Ambjorn:1988tm} the equations of motion were treated in the Bogomolny limit, $M_Z = M_H$, where $M_H = \sqrt{2\lambda} \, v$ and $M_Z = g v /(2 \cos \theta)$ are the masses of the Higgs and $Z$ bosons, respectively. Here we solve -- partially following Ref.~\cite{MacDowell:1991fw} -- the equations of motion~\eq{eq:EoM1}--\eq{eq:EoM6} for arbitrary mass of the Higgs in the region $B \geqslant B_c$ near the phase transition point, $|B - B_c| \ll B_c$. The latter condition implies that the quantity 
\beqn
\epsilon =\frac{|W|}{M_W} \ll 1\,,
\label{eq:epsilon}
\eeqn
can serve as a small expansion parameter.

\begin{figure}[!thb]
\begin{center}
\includegraphics[scale=0.37,clip=false]{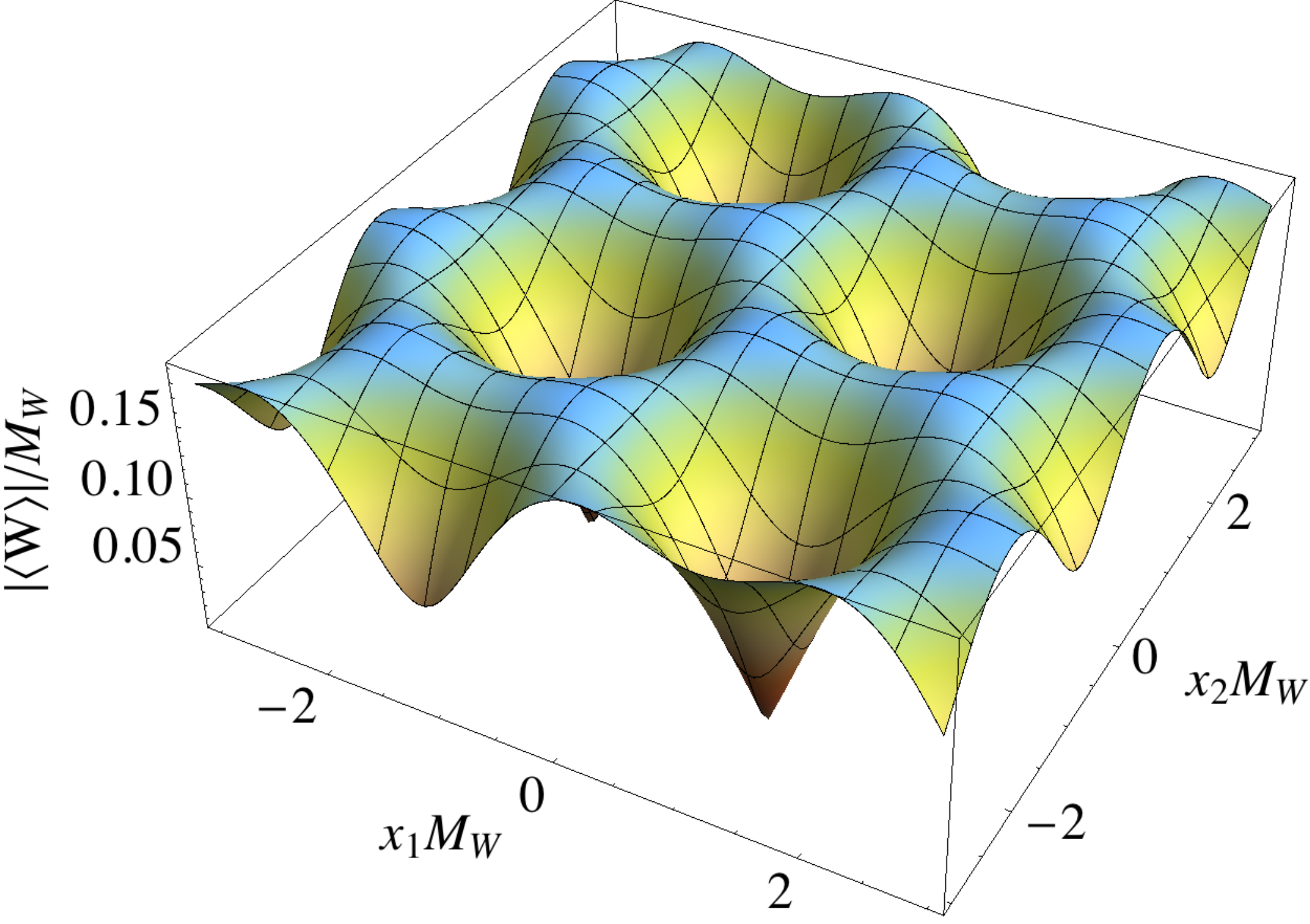}  \\[2mm] 
\hskip 6mm (a) \\[3mm]
\includegraphics[scale=0.37,clip=false]{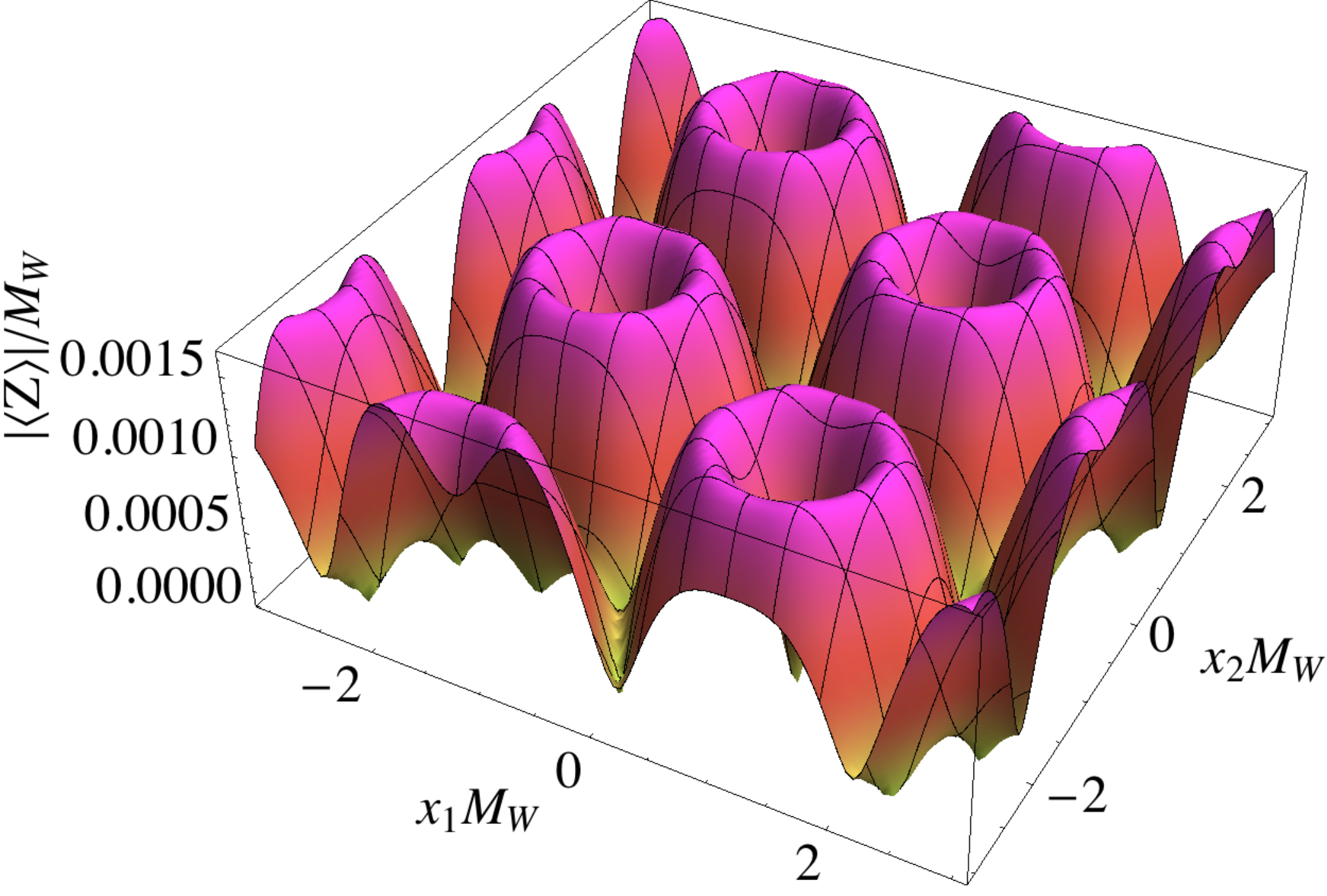}   \\[2mm] 
\hskip 6mm (b) \\[3mm]
\includegraphics[scale=0.36,clip=false]{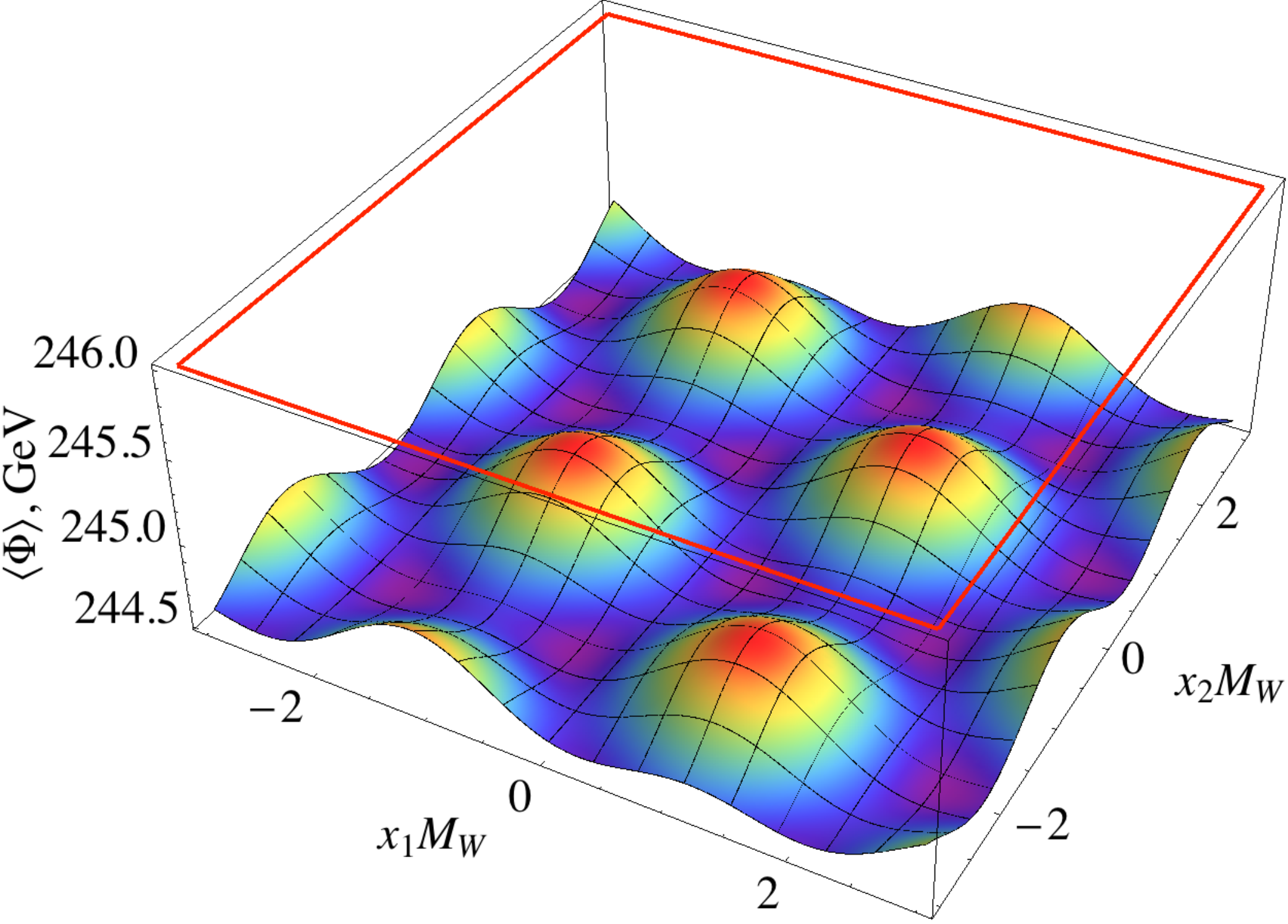} \\[2mm] 
\hskip 6mm (c)
\end{center}
\caption{(a) The superconducting  $W$ condensate~\eq{eq:phi:z:GL}; (b) the superfluid $Z$ condensate~\eq{eq:Z}; (c) the Higgs expectation value~\eq{eq:phi} as a function of the transverse plane coordinates $x_{1}$ and $x_{2}$ at the physical Higgs mass $M_{H} = 125 \, \mbox{GeV}$~\cite{ref:Higgs:mass} in the background magnetic field $B=1.01 B^{\EW}_{c}$ directed along the $x_{3}$ axis. The red line in figure (c) corresponds to the standard (coordinate-independent) Higgs expectation value, $\phi = v/\sqrt{2}$, at zero magnetic field $B = 0$.} 
\label{fig:3d:a}
\end{figure}

\begin{figure*}[!ht]
\begin{center}
\begin{tabular}{cc}
\includegraphics[scale=0.48,clip=false]{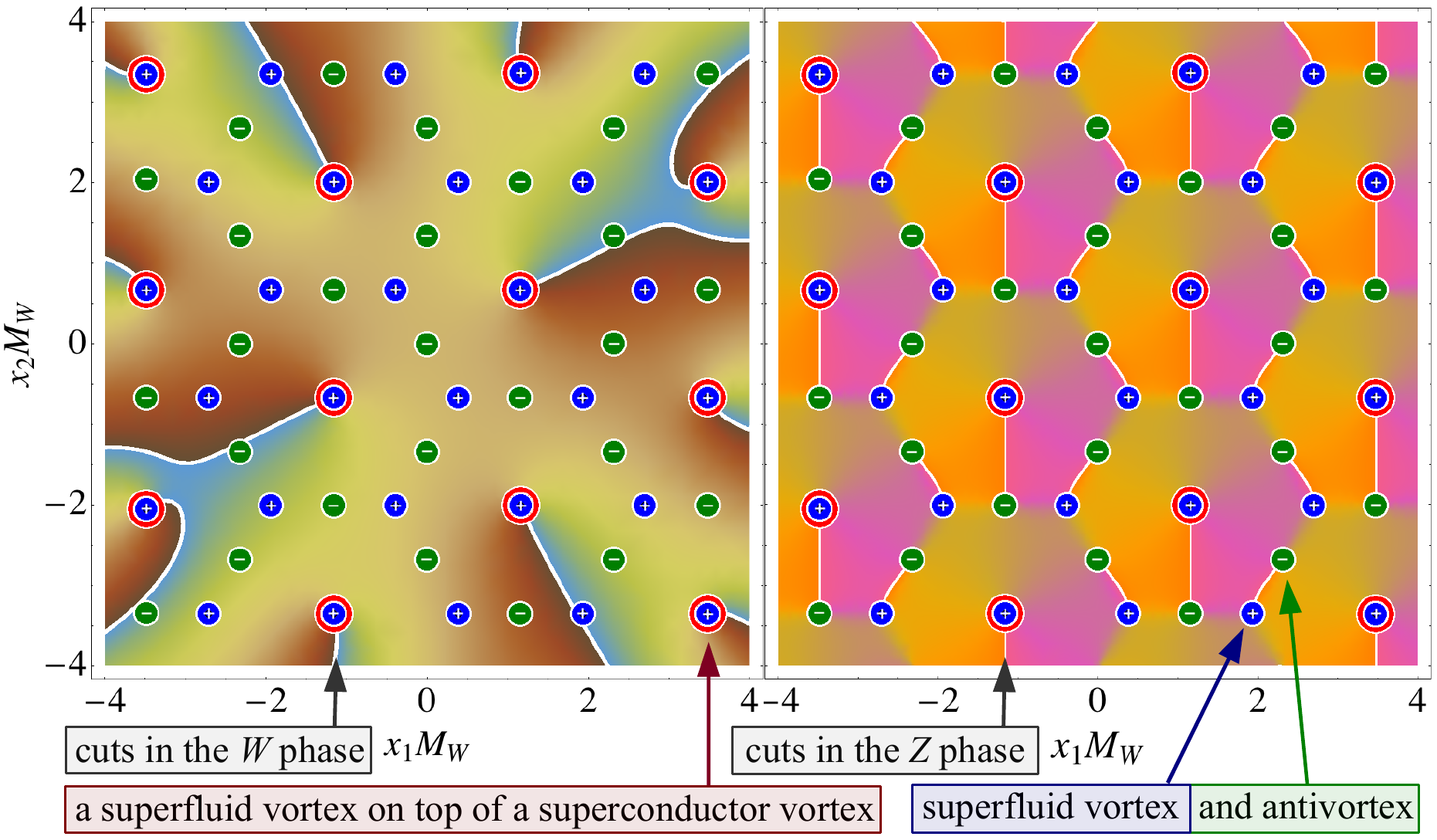} & \qquad
\includegraphics[scale=0.375,clip=false]{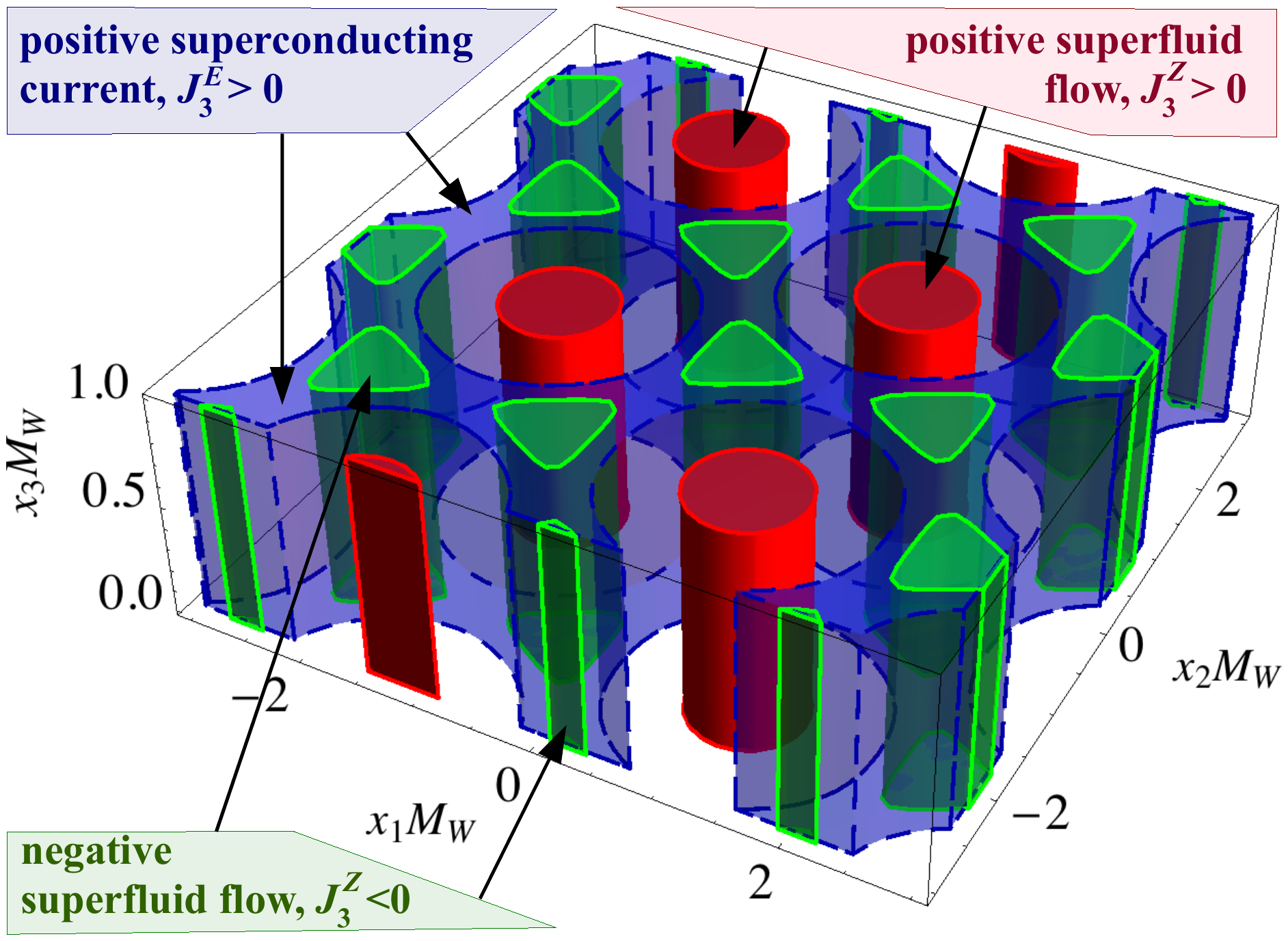} \\[3mm]
\hskip 5mm {\large (a)} & \hskip 10mm  {\large (b)} 
\end{tabular}
\end{center}
\caption{Kaleidoscopic ground state. Figure~(a): The density plots of (the left panel) the phases of the $W$ condensate~\eq{eq:phi:z:GL}; and (right panel) the $Z$ condensate~\eq{eq:Z} in the transversal $(x_{1},x_{2})$ plane at $B = 1.01 \, B^{\EW}_{c}$. The end points of the cuts in the phases (shown by the circles) are the superconductor vortices and superfluid (anti)vortices, respectively; (b) the three-dimensional regions of space predominantly occupied by the superconducting electric current $J^{E}_{3}$ of the $W$ bosons and the superfluid neutral flows $J^{Z}_{3}$ of the $Z$ bosons generated by a weak test electric field $E_{\ext} > 0$ parallel to the strong magnetic field $B = 1.01 \, B^{\EW}_{c}$.}
\label{fig:arguments}
\end{figure*}

\subsection{Vector-meson condensates and vortices}

The combination of the two equations of motion, \eq{eq:EoM1}  and \eq{eq:EoM2}, with the requirement of the minimization of the energy density~\eq{eq:E}, lead us to a simple Abrikosov equation,
\beqn
\mathcal{\bar D}W\approx \mathfrak{\bar D}W = 0\,,
\label{eq:DW}
\eeqn
which is valid up to corrections of the order of $O(\epsilon^2)$. This equation has nontrivial periodic solutions known as Abrikosov lattices. Following Abrikosov~\cite{Abrikosov:1956sx}, we choose a general solution of Eq.~\eq{eq:DW} as a sum over lowest Landau levels:
			\beqn
			W(z) = \sum_{n \in \Z} C_n \, e^{ - \frac{\pi}{2} (|z|^2 + {\bar z}^2) - \pi \nu^2 n^2 + 2 \pi \nu n {\bar z}}\,, 
			\label{eq:phi:z:GL}
			\eeqn
		where $L_B = \sqrt{2 \pi/ (e B)}$ is the magnetic length and $\nu$ is an arbitrary real-valued parameter. In order to ensure a regular structure of the lattice, the complex coefficients $C_n$ are usually chosen in a periodic manner: $C_{n+N} = C_n$, where $N = 1,2, \dots$ is an integer number which has to be chosen using energy-minimization arguments.
	
		The solution with $N = 1$ and $\nu = 1$ defines the square lattice of the original Abrikosov's solution~\cite{Abrikosov:1956sx}. However, the global energy minimum is reached for the equilateral triangular lattice at $N=2$ with $C_1 = \pm i C_0$ and $ \nu = \sqrt[4]{3}/\sqrt{2} \approx 0.9306$ in agreement with earlier studies devoted to the $W$ condensation~\cite{Ambjorn:1988tm,MacDowell:1991fw}. The energy density~\eq{eq:E} is then minimized numerically with respect to the value of $C_0$ for fixed values of the magnetic field $B$ and the Higgs mass $M_H$. This procedure allows us to determine the $W$ condensate~\eq{eq:phi:z:GL} and other interesting quantities.

At $B \geqslant B^{\EW}_{c}$ the condensation of the $W$ bosons, Fig.~\ref{fig:energy}(a), makes the energy density smaller compared to its value in the trivial ground state, Fig.~\ref{fig:energy}(b). Thus, the $W$-boson condensation is an energetically favorable state. Notice that the heavier the Higgs boson the weaker the effect of the magnetic field on the $W$ condensate.
		
Equations~\eq{eq:DW} and \eq{eq:EoM3} imply that the magnetic field $B = B(z) \equiv B(x_1,x_2)$ is related to the $W$ condensate as follows:
\beqn
\partial\bigl(B - e \vert W\vert^2 / 2 \bigr) = 0\,,
\label{eq:B:eq}
\eeqn
This relation is valid up to $O(\epsilon^2)$ terms. The solution of Eq.~\eq{eq:B:eq} carrying a finite energy per unit cell $\cal A$ of the Abrikosov lattice is 
\beqn
B(z) = B_{\ext} + \frac{e}{2}\vert W(z)\vert^2 -\frac{e}{2}\frac{1}{\mathrm{Area} (\mathcal{A})}\int_\mathcal{A} \! d z d {\bar z}\, \vert W\vert^2, \qquad
\label{eq:Bz}
\eeqn
where the integration constant in the last term (given by the integral over the unit lattice cell $\cal A$) guarantees the conservation of the magnetic flux, 
$$\int_\mathcal{A} \! d z d {\bar z}\, B(z) = {\mathrm{Area} (\mathcal{A})} \cdot B_\ext\,.$$
Thus, the magnetic field~\eq{eq:Bz} becomes transversally non\-uniform due to the backreaction of the inhomogeneous $W$ condensate~\eq{eq:phi:z:GL}.

Using the solution~\eq{eq:Bz} for the magnetic field $B \equiv F_{12}$, one can solve Eqs.~\eq{eq:EoM4} and \eq{eq:EoM5} and obtain the following nonlocal expressions for the $Z$ and Higgs condensates, respectively:
			\beqn
			Z \equiv Z_{1} + i Z_{2}&=&  -i\frac{ g \cos \theta}{2} \frac{\partial_{1}\,+\, i \partial_{2}}{- \Delta + M^2_Z}  |W|^2\!,\,
			\label{eq:Z} \\
			\phi &=& \frac{ v}{\sqrt{2}}\left(1 -   \frac{g^2}{4}\frac{1}{- \Delta + M_H^2} \vert W\vert^2\right)\,. \quad
			\label{eq:phi}
			\eeqn
Here $\Delta \equiv {\bar \partial} \partial = \partial_{1}^{2} + \partial_{2}^{2}$ is the two-dimensional Laplacian in the transverse plane. The remaining equation~\eq{eq:EoM6} is satisfied automatically up to $O(\epsilon^2)$.
		
In the ground state at $B > B^\EW_c$, the $W$ condensate~\eq{eq:phi:z:GL}, the $Z$ condensate~\eq{eq:Z}  and the Higgs condensate~\eq{eq:phi} are functions of the transversal coordinates $x_{1}$ and $x_{2}$, as visualized in Figs.~\ref{fig:3d:a}(a), (b) and (c), respectively. 
The expectation value of the Higgs field falls down as the magnetic field rises, with a slope which becomes weaker as the Higgs mass increases, Fig.~\ref{fig:3d:a}(c).

It is known that the ground state of the vacuum at $B > B^{\EW}_{c}$ is an equilateral triangular lattice of the vortex defects in the $W$ field~\cite{Ambjorn:1988tm,MacDowell:1991fw} (we call these vortices  the ``superconductor vortices''). At the vortex positions the field $W \propto W^{-}_{1} + i W^{-}_{2}$ vanishes, Fig.~\ref{fig:3d:a}(a), and its phase, $\mathrm{arg} (W)$, winds around each vortex position. We find that the ground state has a much more complicated structure in the neutral sector: the state has an equilateral triangular lattice of the ``superfluid'' vortices, characterized by the vanishing field $Z \equiv Z_{1} + i Z_{2}$ field, Fig.~\ref{fig:3d:a}(b), and by winding numbers in its phase\footnote{The superconductor and superfluid vortices, which are discussed in this paper, should be distinguished from the existing $W$-- and $Z$-- electroweak vortex solutions~\cite{Achucarro:1999it}, including known solutions which carry electric currents along vortex cores~\cite{Volkov:2006ug}.}. The combined ``kaleidoscopic'' pattern of the vortex lattices, superimposed on the density plots of the phases of the superconducting, $W$ and the superfluid $Z$ fields, are shown in Fig.~\ref{fig:arguments}(a). Notice that certain superfluid vortices are located at the superconductor vortices.

\section{Nondissipative transport}
\label{sec:transport}

\begin{figure}[!thb]
\begin{center}
\begin{tabular}{ccc}
\includegraphics[scale=0.34,clip=false]{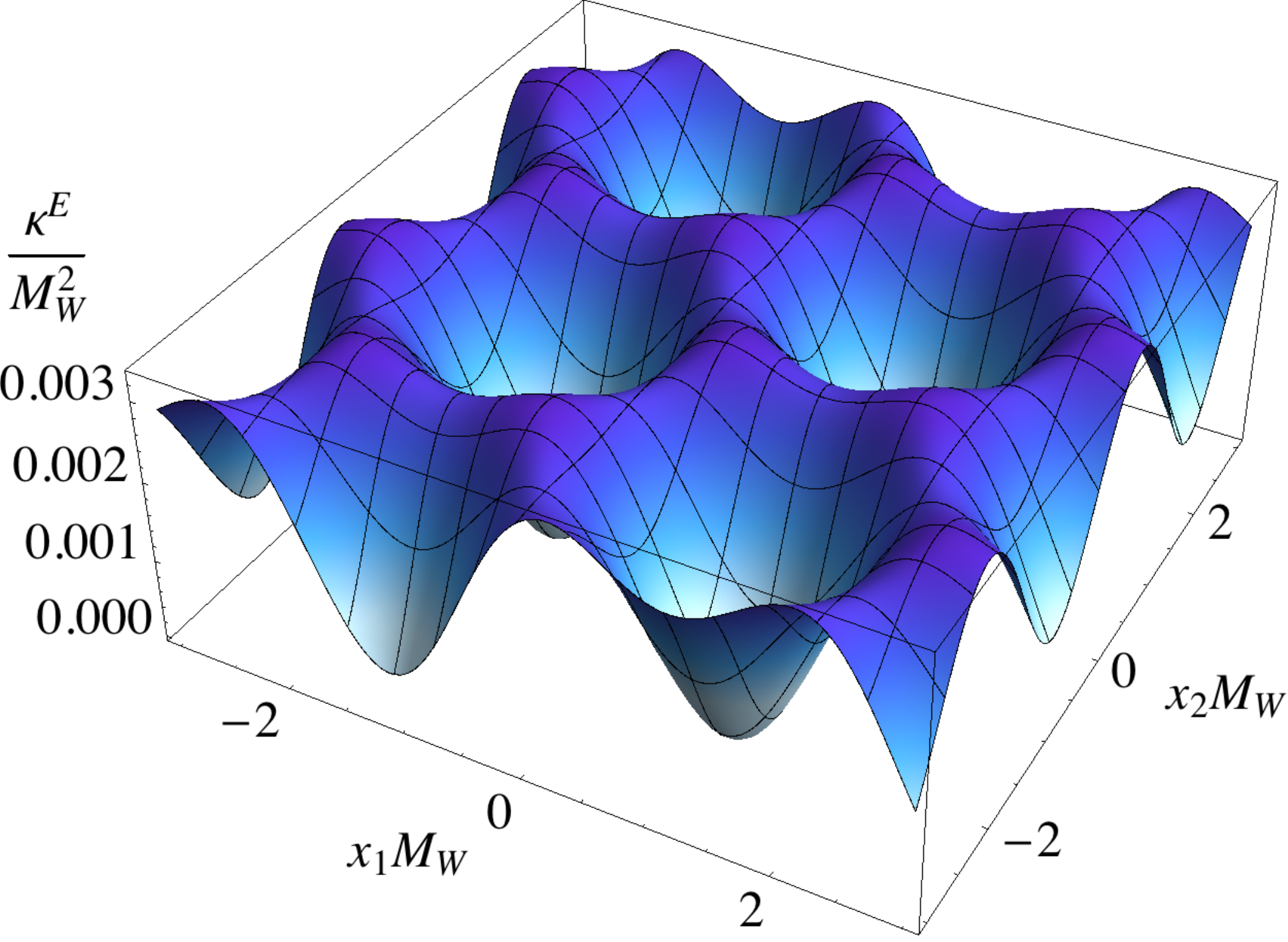}  \\[2mm] 
\hskip 6mm (a) \\[3mm]
\includegraphics[scale=0.34,clip=false]{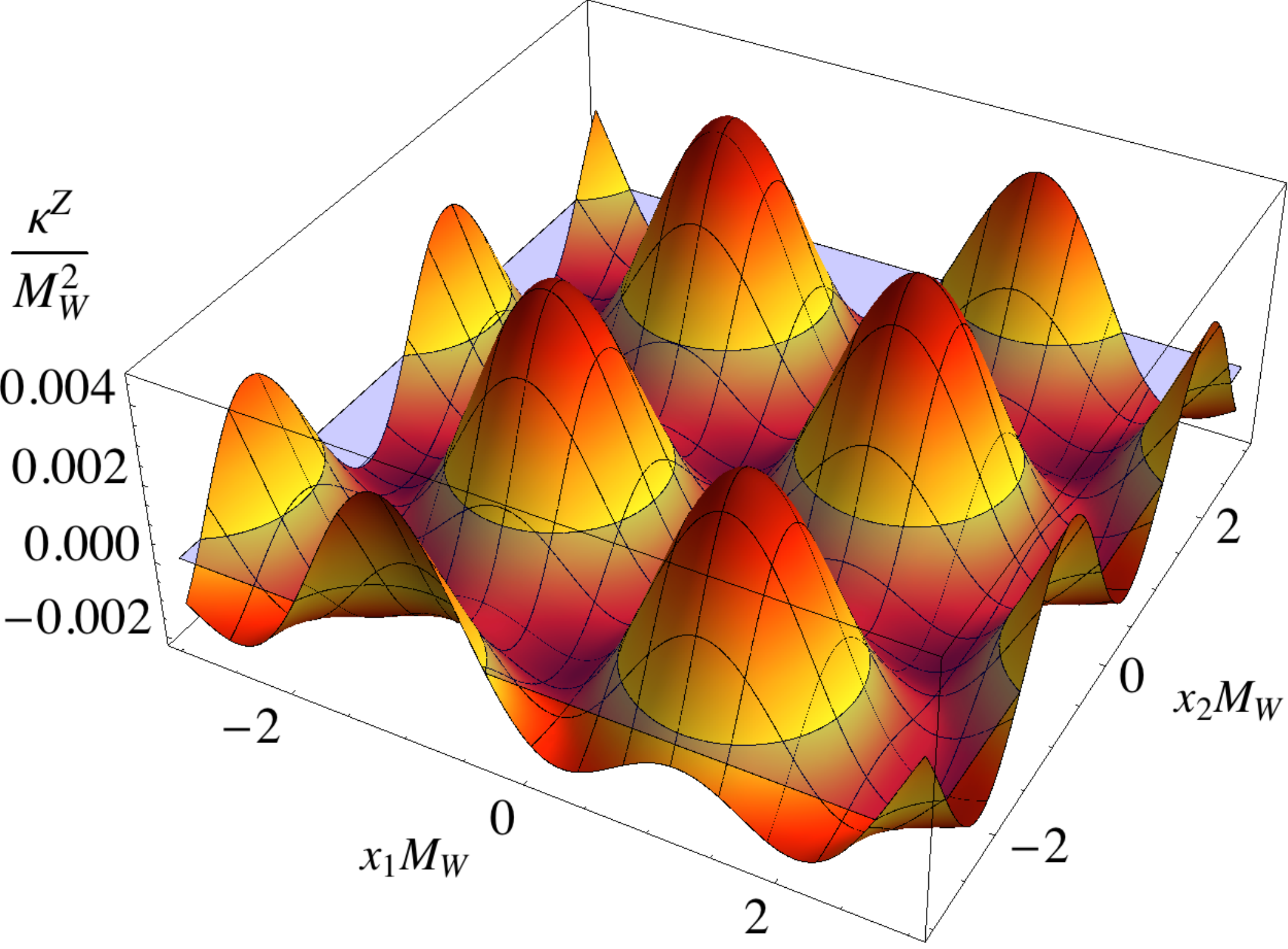}  \\[2mm] 
\hskip 6mm (b) \\[3mm]
\includegraphics[scale=0.5,clip=false]{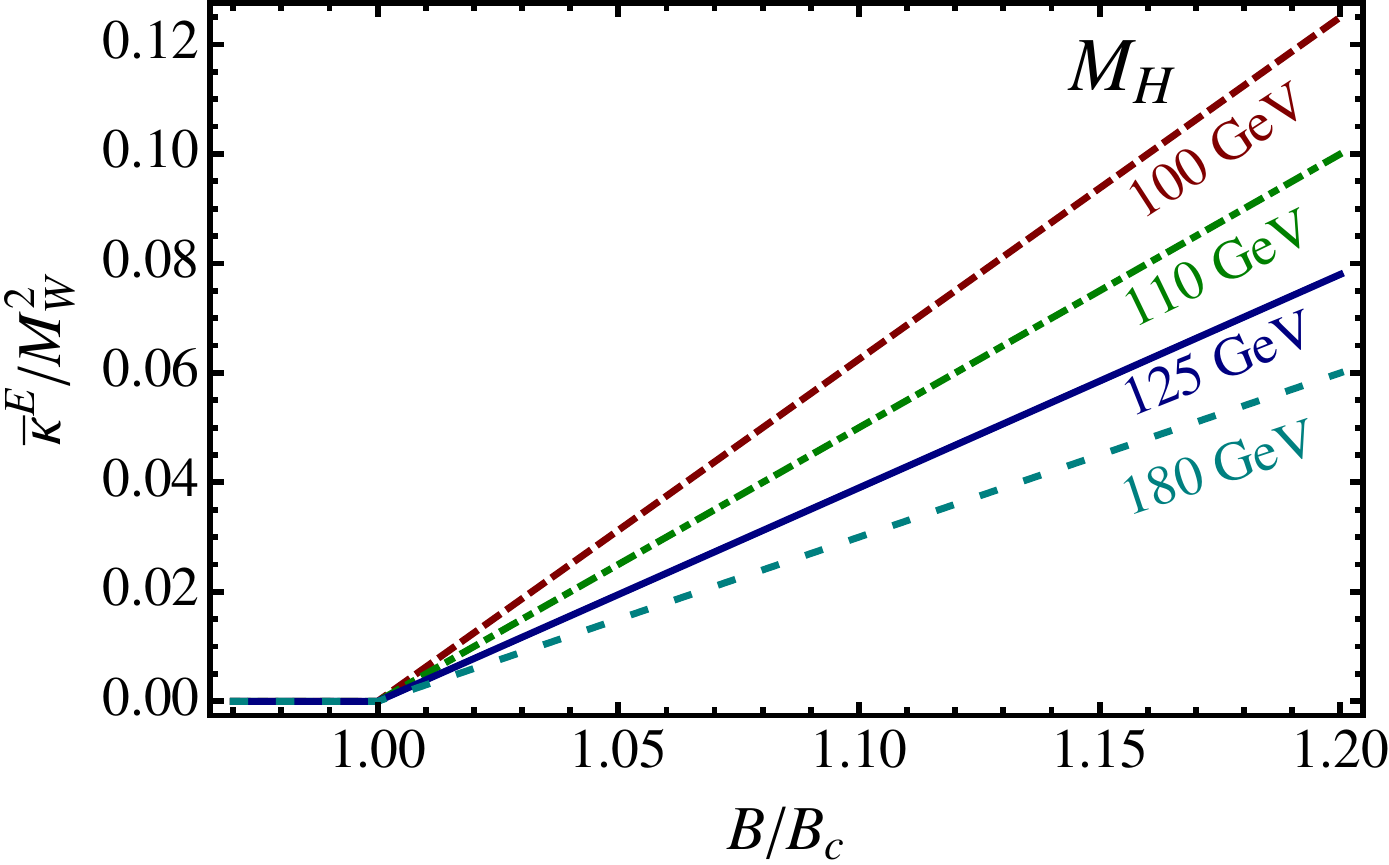} \\[2mm] 
\hskip 6mm (c)
\end{tabular}
\end{center}
\caption{(a) The superconducting~\eq{eq:kappaE} and (b) superfluid~\eq{eq:kappaZ} transport coefficients as the functions of the transverse plane coordinates $x_{1}$ and $x_{2}$ at the physical Higgs mass $M_{H} = 125 \, \mbox{GeV}$ in the background magnetic field $B=1.01 B^{\EW}_{c}$ directed along the $x_{3}$ axis. (c) The cell-averaged superconductivity transport coefficient~\eq{eq:kappaE} {\it vs.} the magnetic field $B$ at fixed values of the Higgs masses. The cell-averaged superfluidity coefficient is always zero.} 
\label{fig:3d:b}
\end{figure}

We point out that the ground state of the vacuum at $B > B^{\EW}_{c}$ is a ``tandem'' phase which is, simultaneously, an electromagnetic superconductor and a neutral superfluid. Indeed, introducing an  infinitesimally weak test electric field $E^{\ext}$ one can prove -- with the use of Eq.~\eq{eq:classical} -- the following transport laws for the electromagnetic and neutral $Z$-boson currents,
\beqn
J^{E}_{\mu} & = & \partial^\nu F_{\nu\mu}\propto \frac{\delta \cL }{\delta A^{\mu}}\,, \\
J^{Z}_{\mu} & = & \partial^\nu Z_{\nu\mu}\propto \frac{\delta \cL }{ \delta Z^{\mu}}\,,
\eeqn
respectively:
			\beqn
			\partial_{[0} J_{3]}^{E}(x) & = & - \kappa^{E}(x_{1},x_{2}) \cdot E^{\ext}_{3}\,, \qquad \partial_{[0} J_{i]}^{E} = 0\,, \quad
			\label{eq:London:E}\\
			\partial_{[0} J_{3]}^{Z}(x) & = & - \kappa^{Z}(x_{1},x_{2}) \cdot E^{\ext}_{3}\,,      \qquad \partial_{[0} J_{i]}^{Z} = 0\,, \quad
			\label{eq:London:Z}
			\eeqn
		where $i=1,2$. The transport parameters for the electromagnetic $\kappa^{E}$ and neutral $\kappa^{Z}$ currents,
			\beqn
			\kappa^{E}(x_{1},x_{2}) & = & e^{2} |W|^{2} (x_{1},x_{2})\,, 
			\label{eq:kappaE} \\
			\kappa^{Z}(x_{1},x_{2}) & = & -e^{2} \cot \theta \frac{\Delta}{-\Delta+M^{2}_{Z}} |W|^{2} (x_{1},x_{2})\,.
			\label{eq:kappaZ} 
			\eeqn
are the functions of the transverse coordinates $x_{1}$ and $x_{2}$. These transport coefficients are shown in Figs.~\ref{fig:3d:b}(a) and (b), respectively.

Equation \eq{eq:London:E} implies anisotropic superconductivity of the ground state at $B > B_{c}^{\EW}$ similarly to an analogous phenomenon in QCD~\cite{Chernodub:2010qx,Chernodub:2011mc}: a weak electric field introduces a resistance-free growth of electric current which continues streaming after the field is switched off. Equation~\eq{eq:London:Z} implies an anisotropic superfluidity of the neutral $Z$ currents, and it illustrates a very unusual physical effect: an external electric field induces a current of neutral particles which are flowing frictionlessly along the magnetic field axis. 
		
From the point of view of the electric conductivity properties, a ground state of the vacuum can either be a superconductor or an insulator due to Lorentz symmetry (indeed, a dissipative behavior, like in the Ohm's Law, is inconsistent with the Lorentz symmetry of the vacuum). Thus, the absence of the electric resistance (and vanishing shear and bulk viscosities) in the $B > B_{c}^{\EW}$ phase are protected by a remnant Lorentz symmetry in the $(x_{0},x_{3})$ plane. Similar Lorentz-protection arguments apply to the superfluid property as well.
		
The  superconductivity coefficient~\eq{eq:kappaE}, averaged over the transversal $(x_{1},x_{2})$ plane, 
\beqn
{\bar \kappa}^E = \frac{1}{\mathrm{Area} (\mathcal{A})}\int_\mathcal{A} \! d x_1 \, d x_2 \, {\kappa}^E(x_1,x_2)\,, 
\eeqn		
is a linearly growing function of the magnetic field $B$, Fig.~\ref{fig:energy}(d), at $B > B_{c}^{\EW}$. The superfluid coefficient~\eq{eq:kappaZ} is a sign-changing function, Fig.~\ref{fig:3d:b}(b), of the transversal coordinates $x_{1,2}$ which has a vanishing mean value if averaged over the transversal plane (${\bar \kappa}^Z \equiv 0$). 

Thus, we conclude that a weak external electric field $E^{\ext}_{3}$ applied along the magnetic field in the condensed phase gives rise to 
\begin{itemize}
\item[(i)] a growing nonzero net electric current along the magnetic field axis, and
\item[(ii)] a neutral superfluid inhomogeneous flow in both directions with vanishing net current. 
\end{itemize}

The spatial distribution of the electric and neutral currents flowing along the magnetic field axis can be read off from the corresponding superconducting coefficients in Figs.~\ref{fig:3d:b}(a) and (b), respectively. The distribution of the currents in the transverse place is visualized in Fig.~\ref{fig:arguments}(b).		

Notice that the transverse electric field $E^{\ext}_{1,2}$ induces neither superconducting nor superfluid currents.

\section{Conclusion}

We have shown for the first time that the electroweak sector of the vacuum exhibits superconducting and superfluid properties due to the magnetic-field-induced condensation of, respectively, $W$ and $Z$ bosons provided the magnetic field exceeds the critical value~\eq{eq:eBc}. The superconductor-superfluid phase is characterized by the anisotropic and inhomogeneous ground state. Both charged and neutral currents may propagate nondissipatively only along the direction of the magnetic field. In the transverse directions the ground state has an unusual ``kaleidoscopic'' structure made of a hexagonal lattice of superfluid vortices superimposed on an equilateral triangular (hexagonal) lattice of superconductor vortices. Thus, in a strong enough magnetic field the electroweak sector of the quantum vacuum enters a superconductor-superfluid phase.

\acknowledgments

The work of MNC was supported by Grant No. ANR-10-JCJC-0408 HYPERMAG (France).


\begin{thebibliography}{99}
\bibitem{Fukushima:2008xe}
  K.~Fukushima, D.~E.~Kharzeev and H.~J.~Warringa,
  Phys.\ Rev.\  D {\bf 78}, 074033 (2008).

\bibitem{Kharzeev:2004ey}
  D.~Kharzeev,
  Phys.\ Lett.\  B {\bf 633}, 260 (2006).

\bibitem{Vilenkin:1980fu}
  A.~Vilenkin,
  Phys.\ Rev.\  D {\bf 22}, 3080 (1980).

\bibitem{Skokov:2009qp}
  V.~Skokov, A.~Y.~Illarionov and V.~Toneev,
  Int.\ J.\ Mod.\ Phys.\  A {\bf 24}, 5925 (2009).

\bibitem{Deng:2012}
  W.~T.~Deng and X.~G.~Huang,
  Phys.\ Rev.\ C {\bf 85}, 044907 (2012) [arXiv:1201.5108 [hep-ph]];
  A.~Bzdak and V.~Skokov,
  Phys.\ Lett.\ B {\bf 710}, 171 (2012)
  [arXiv:1111.1949 [hep-ph]].
   
\bibitem{Ferrer:2005vd} 
  E.~J.~Ferrer, V.~de la Incera and C.~Manuel,
  Phys.\ Rev.\ Lett.\  {\bf 95}, 152002 (2005);
  S.~Fayazbakhsh and N.~Sadooghi,
  Phys.\ Rev.\ D {\bf 82}, 045010 (2010);
  A.~Rabhi, P.~K.~Panda and C.~Providencia,
  Phys.\ Rev.\ C {\bf 84}, 035803 (2011);
  A.~A.~Isayev and J.~Yang,
{\it ibid.}
  {\bf 84}, 065802 (2011).

\bibitem{Adler:1971wn} 
  S.~L.~Adler,
  Annals Phys.\  {\bf 67}, 599 (1971).

\bibitem{Klimenko:1991he}
  S.~P.~Klevansky, R.~H.~Lemmer,
  Phys.\ Rev.\ D {\bf 39}, 3478 (1989);
  H.~Suganuma, T.~Tatsumi,
  Annals Phys.\  {\bf 208}, 470 (1991);
  K.~G.~Klimenko,
  Z.\ Phys.\  C {\bf 54}, 323 (1992);
  V.~P.~Gusynin, V.~A.~Miransky and I.~A.~Shovkovy,
  Phys.\ Rev.\ Lett.\  {\bf 73}, 3499 (1994);
  V.~P.~Gusynin, V.~A.~Miransky and I.~A.~Shovkovy,
  Nucl.\ Phys.\  B {\bf 462}, 249 (1996).

\bibitem{Grasso:2000wj}
  D.~Grasso, H.R.~Rubinstein,
  Phys.\ Rept.\  {\bf 348}, 163 (2001).

\bibitem{Chernodub:2010qx}
  M.~N.~Chernodub,
  Phys.\ Rev.\  D {\bf 82}, 085011 (2010).

\bibitem{Chernodub:2011mc}
  M.~N.~Chernodub,
  Phys.\ Rev.\ Lett.\  {\bf 106}, 142003 (2011).

\bibitem{Chernodub:2011gs}
  M.~N.~Chernodub, J.~Van Doorsselaere and H.~Verschelde,
  Phys.\ Rev.\  D {\bf 85}, 045002 (2012).
 
\bibitem{Smolyaninov:2011wc} 
  I.~I.~Smolyaninov,
  Phys.\ Rev.\ Lett.\  {\bf 107}, 253903 (2011).

\bibitem{ref:second:transition:1}
  A.~Salam and J.~A.~Strathdee,
  Nucl.\ Phys.\ B {\bf 90}, 203 (1975);
  A.~D.~Linde,
  Phys.\ Lett.\ B {\bf 62}, 435 (1976).

\bibitem{ref:second:transition:2}
   J.~Van Doorsselaere,
  Phys. Rev. D {\bf 88}, 025013 (2013) [arXiv:1206.6205 [hep-ph]].

\bibitem{Ambjorn:1988tm} 
  J.~Ambjorn and P.~Olesen,
  Nucl.\ Phys.\ B {\bf 315}, 606 (1989);
  Phys.\ Lett.\ B {\bf 218}, 67 (1989);
  Int.\ J.\ Mod.\ Phys.\ A {\bf 5}, 4525 (1990).

\bibitem{MacDowell:1991fw} 
  S.~W.~MacDowell and O.~Tornkvist,
  Phys.\ Rev.\ D {\bf 45}, 3833 (1992);
  O.~Tornkvist,
  hep-ph/9204235.

\bibitem{Abrikosov:1956sx} 
  A.~A.~Abrikosov,
  Sov.\ Phys.\ JETP {\bf 5}, 1174 (1957).

\bibitem{ref:Higgs:mass} 
  G.~Aad {\it et al.}  [ATLAS Collaboration],
  Phys.\ Lett.\ B {\bf 716}, 1 (2012)
  [arXiv:1207.7214 [hep-ex]];
  S.~Chatrchyan {\it et al.}  [CMS Collaboration],
  Phys.\ Lett.\ B {\bf 716}, 30 (2012)
  [arXiv:1207.7235 [hep-ex]].

\bibitem{Achucarro:1999it} 
See, e.g., a review in  A.~Achucarro and T.~Vachaspati,
  Phys.\ Rept.\  {\bf 327}, 347 (2000).

\bibitem{Volkov:2006ug} 
  M.~S.~Volkov,
  Phys.\ Lett.\ B {\bf 644}, 203 (2007);
  J.~Garaud and M.~S.~Volkov,
  Nucl.\ Phys.\ B {\bf 826}, 174 (2010).

\end{thebibliography}
\end{document}